\documentclass[prx,aps,showpacs,footinbib,twocolumn]{revtex4-1}
\usepackage{graphicx}
\usepackage{float}
\usepackage{placeins}
\usepackage{gensymb}
\usepackage{bm}
\usepackage{amssymb}
\usepackage{amsmath}
\usepackage{color}
\usepackage{float}
\usepackage{ctable}
\usepackage{braket}
\usepackage{epstopdf}
\usepackage{mathtools}
\usepackage{mathrsfs}
\usepackage{natbib}
\usepackage[resetlabels]{multibib}
\newcites{som}{appendix}
\usepackage[colorlinks, linkcolor=blue, citecolor=blue, urlcolor=blue,
breaklinks=true]{hyperref}

\allowdisplaybreaks

\begin{document}

\title{Synchronization of Interacting Quantum Dipoles}

\author{B.~Zhu$^{1}$}
\author{J.~Schachenmayer$^1$}
\author{M.~Xu$^1$}
\author{F.~Herrera$^{2}$}
\author{J.~G.~Restrepo$^{3}$}
\author{M.~J.~Holland$^{1}$}
\email{murray.holland@colorado.edu}
\author{A.~M.~Rey$^{1}$}
 \email{arey@jilau1.colorado.edu}

\affiliation{$^{1}$JILA, NIST, Department of Physics, University of Colorado, 440 UCB, Boulder, CO 80309, USA}
\affiliation{$^{2}$ Department of Physics, Universidad de Santiago de Chile, USACH, Casilla 307 Correo 2 Santiago, Chile}
\affiliation{$^{3}$ Department of Applied Mathematics, University of Colorado, Boulder, Colorado, 80309, USA}

\date{\today}

\begin{abstract}
 Macroscopic ensembles of radiating  dipoles are ubiquitous in the physical and natural sciences.  In the classical limit the dipoles can be described as damped-driven oscillators, which are able to  spontaneously synchronize and collectively lock their phases in the presence of nonlinear coupling. Here we  investigate the corresponding phenomenon
 with  arrays of quantized two-level systems coupled  via long-range and anisotropic dipolar interactions.   Our calculations demonstrate that by incoherently driving  dense packed arrays of  strongly interacting  dipoles,  the dipoles can overcome the decoherence induced by quantum fluctuations and inhomogeneous
  coupling  and reach a  synchronized steady-state  characterized by a macroscopic phase coherence. This
  steady-state bears much similarity to that observed in classical
  systems, and yet also exhibits genuine quantum properties such as
  quantum correlations and quantum phase diffusion (reminiscent of
  lasing). Our predictions could be relevant for  the development of  better atomic clocks and a variety of  noise tolerant quantum devices.
\end{abstract}

\pacs{ 05.45.Xt, 42.50.Lc, 37.10.Jk,  64.60.Ht}

\maketitle
\section{introduction}
Arrays of synchronized oscillators ~\cite{pikovsky} are ubiquitous in
biological~\cite{cellularscience,syncthalamus}, physical
~\cite{precisionpre} and engineering ~\cite{powerstation} systems and
are a resource for technological advances~\cite{PM}. Although there
has been significant progress in the study of synchronization in
classical systems~\cite{syncnetworks}, the understanding of the same phenomena in the quantum realm
remains limited. A major obstacle so far is the general problem of
the exponential scaling of the Hilbert space with system size which
makes calculations dealing with quantum arrays very challenging. In fact, current investigations
have been limited to the exact  treatment of arrays of a small
number of coupled quantum oscillators
\cite{Walter2014,Shim2007,Zhang2012,Matheny2014,Giorgi2012,Giorgi2013,Qiu2014,Ameri2014,Hush2014,syncprl2014,Manzano2013}, and large ensembles  at the mean field level or
by including quantum corrections perturbatively \cite{Ludwig2013,Lee2014,Lee2013}. Highly symmetric situations with collective coupling mediated,
for example, by a cavity mode
\cite{Bagheri013,Xu2013b,Mari2013}, have also been studied.

Ensembles of radiating dipoles are a natural platform to study quantum synchronization, where coherence can be  generated from an incoherent source. One might regard   laser systems, where radiation is amplified by the stimulated emission of photons, as a prototypical  example. However,  lasing   is fundamentally a distinct phenomenon  from quantum synchronization. This can be seen from the fact that lasing is possible even in the absence of  coupling between the atomic dipoles, as  is clear in the single atom laser \cite{singatom}, or in
atomic beam lasers where only one atom is present in the cavity at any
given time. A more relevant situation is the  quantum synchronization that takes place in the context of
superradiance~\cite{haroche1982}. It has recently been understood that, in contrast to
lasers, steady-state superradiance   can  produce spectrally
pure light~\cite{haroche1982, suplaser,dmeiser2} without stimulated emission.
So far this has been demonstrated using a cavity mode  as a communication channel that spatially selects an optical mode and enhances  the coupling (through the cavity
finesse). A more generic and  relevant  scenario,  with great potential and applicability,  is the emergence of  spontaneous macroscopic quantum synchronization  in radiating dipole arrays without a cavity but  naturally coupled by the  intrinsic anisotropic and long-range dipolar interactions. This is the situation considered in this paper.

\begin{figure*}[t]
\centering
  \includegraphics[width=1.7\columnwidth]{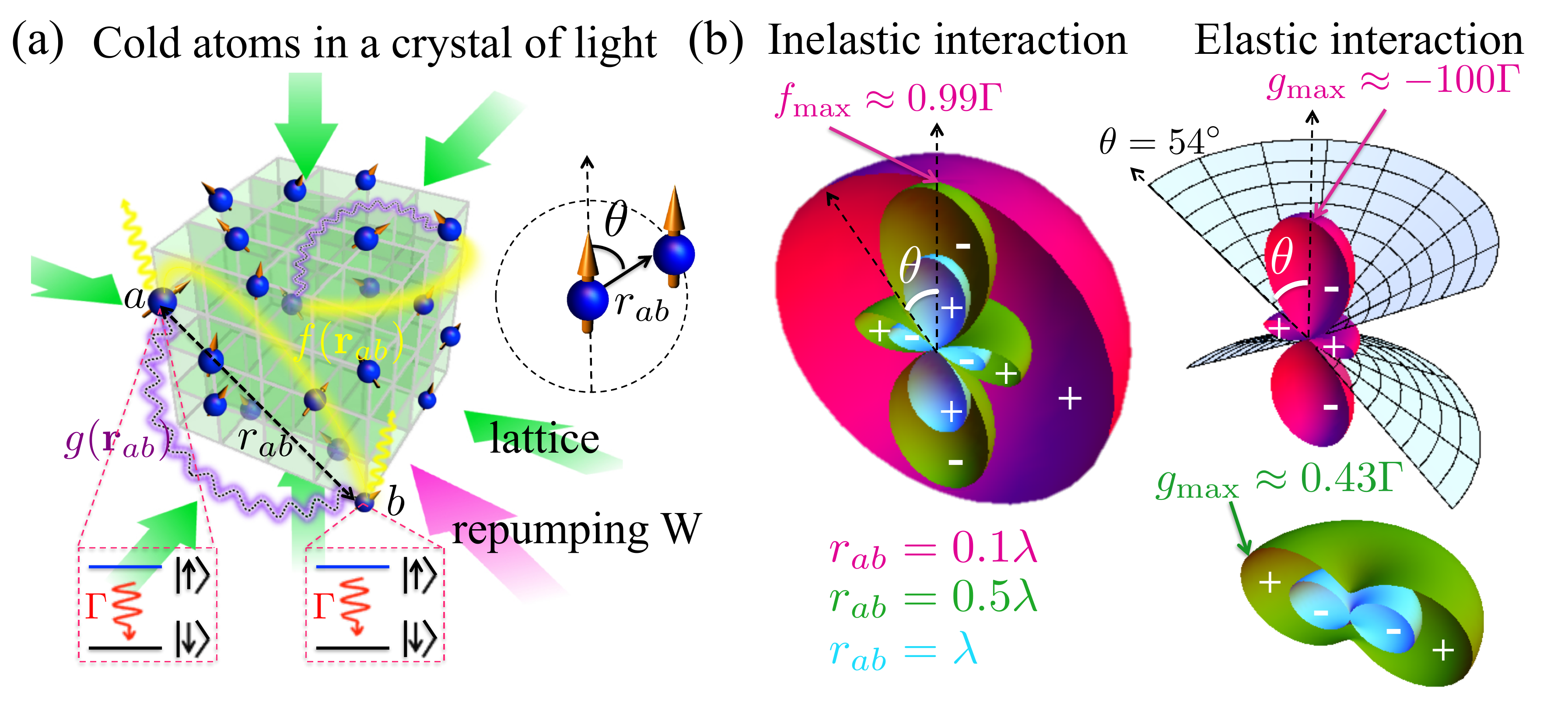}\caption{Arrays
    of quantum dipoles spontaneously emit and absorb photons at rate
    $\Gamma$. The photons mediate dipolar interactions between dipoles
    separated by a distance ${\bf r}_{ab}$ with both dissipative,
    $f({\bf r}_{ab})$, and elastic, $g({\bf r}_{ab})$, components.  A
    repumping source at a rate ~$W$ provides energy to maintain the
    oscillations and can be implemented using additional
    internal states that are not shown. (a) A possible implementation using cold atoms in an optical lattice.
     (b) The dipolar couplings $g({\bf r}_{ab})$ and $f({\bf
      r}_{ab})$ exhibit a complex angular distribution as a function
    of $\theta$, the angle between the dipole orientation (determined
    by an external electromagnetic field) and~$|{\bf r}_{ab}|$. The
    maximum value of $f$ and $g$ for fixed $|\bf {r}_{ab}|$ is denoted
    as ${f}_{\rm max}$ and $g_{\rm max}$.  The cone illustrates the
    magic angle, $\theta_m=\arccos(1/\sqrt{3})$.}\label{fig:fig1}
\end{figure*}

Here we demonstrate that in the presence of an incoherent repumping source,
dipole induced cooperative emission can dominate over spatial
inhomogeneities and quantum fluctuations and lead to a
resilient steady-state that exhibits macroscopic quantum phase coherence
and intrinsic quantum correlations.  An iconic example of a macroscopic coherent state is a Bose-Einstein condensate, achieved in  ultra-cold gases at  thermal equilibrium. In our case, however, the macroscopic order is reached  in the  steady-state of an interacting and  driven-dissipative system. Moreover, the cooperative behavior
can be detected by measuring the spectral purity of the emitted
radiation. We note that  in clear distinction to  previous studies
 ~\cite{Hush2014,Ludwig2013,Walter2014,Ameri2014,syncprl2014,Lee2013}, our proposal  does not rely on  an external coherent source or externally  generated nonlinearities to seed the collective phase. In our model synchronization emerges as a spontaneously broken symmetry   driven by incoherent processes in  naturally coupled  dipole arrays. As we show, and somewhat counterintuitively, an incoherent drive  is sufficient to generate phase coherence in these systems.

  Specifically, the systems we consider are dense arrays of frozen quantum dipoles modeled as quantized two-level systems. By dense arrays of frozen dipoles we mean arrays separated
by a distance much closer than the wavelength of the emitted photons
and with motional degrees of freedom evolving at a much slower rate
than their internal dynamics (Fig.~\ref{fig:fig1}). These conditions
can be readily satisfied in a variety of quantum systems found in
atomic, molecular and optical physics (e.g., Rydberg
gases~\cite{vanDitzhuijzen2008,Afrousheh2004,Anderson1998}, alkali
vapors~\cite{Keaveney2012}, alkaline-earth atoms~\cite{Olmos2013}, and
polar molecules~\cite{yan2013}), chemistry (e.g., J-aggregates of dye
molecules~\cite{elasticvsinleastic,superradiantaggregate,spano1989}),
and biology (e.g., light-harvesting
complexes~\cite{quantumcoherence,quantumchemestryplans}). In cold
vapors, one possible way to freeze the motion and tightly trap the
particles is via an optical lattice potential (Fig.~\ref{fig:fig1}). In this case a sub-optical-wavelength transition must be used  in order to reach the tight-packing regime~\cite{yan2013,Olmos2013}.

To fully understand synchronization in the complex dipolar system, we analyze   each of the ingredients that compete and affect synchronization in a step-by-step procedure: the interplay between repumping and collective emission,  inhomogeneity in the coupling constants, quantum correlations,  and the competition between elastic and inelastic interactions.
The paper is organized as follows: In Sec.~II we introduce the system in consideration and the master equation we use to describe the dynamics. In Sec.~III we first provide a simple mean-field description and discuss connections to the classical  Kuramoto model---the iconic model used to describe synchronization in  non-linear coupled oscillators. In Sec.~IV we discuss the phase diagram for the quantum system assuming collective (all-to-all) coupling and compare it with the mean-field solution. For this exactly solvable case we are able to explicitly  quantify the entanglement and correlations present in the steady state. In Sec.~V we study  how  inhomogeneity in the inelastic couplings affects synchronization and focus on the case of  power-law decaying interactions. In Sec.~VI we study the emergence of  quantum synchronization in radiating dipoles taking the full long-range and anisotropic dipolar interactions into account. In Sec.~VII we discuss experimental implementations of our model,  and in Sec.~VIII we provide a conclusion and an outlook.

\section{dipole-dipole interaction and master equation}

In this work we consider arrays of  quantum dipoles with
two accessible levels, which we denote as $\ket{\downarrow}$  and~$\ket{\uparrow}$. The  interactions between  two  dipoles $a$ and $b$ are described by the  functions $g({\bf r}_{ab})$ and $f({\bf r}_{ab})$, which depend on  the dipoles'   separation, $|{\bf r}_{ab}|$,
 and  the angle $\theta$ between  the mean dipole moment and   the  vector joining the dipoles [See Fig.~\ref{fig:fig1}(a)] \cite{Sup}:
\begin{align}
  &g({\bf r}_{ab})\!=\!-\frac{3\Gamma}{2}\Big\{\text{sin}^2\theta
  \frac{\text{cos}\zeta_{ab}}{\zeta_{ab}}\!+\!(3\text{cos}^2\theta\!-\!1)
  [\frac{\text{cos}\zeta_{ab}}{(\zeta_{ab})^3}\!+\!\frac{\text{sin}\zeta_{ab}}
  {(\zeta_{ab})^2}]\Big\}\nonumber \\
  &f({\bf r}_{ab})\!=\!\frac{3\Gamma}{2}\Big\{\text{sin}^2\theta
  \frac{\text{sin}\zeta_{ab}}{\zeta_{ab}}\!+\!(3\text{cos}^2\theta\!-\!1)
  [\frac{\text{sin}\zeta_{ab}}{(\zeta_{ab})^3}\!-\!\frac{\text{cos}\zeta_{ab}}
  {(\zeta_{ab})^2}]\Big\}. \nonumber
\end{align}
Here, $\zeta_{ab}=2\pi |{\bf r}_{ab}|/\lambda$, where $\lambda$ is the characteristic wavelength of the dipole-transition, and $\Gamma=f(0)$ is the  spontaneous photon emission rate from a single dipole. The function $g({\bf r}_{ab})$  describes the elastic dipole-dipole interactions, while $f({\bf r}_{ab})$ gives rise to inelastic collective photon emission. These terms are similar to those  that determine
the radiation of  classical electric dipoles, and the
 dependence on $|{\bf r}_{ab}|$ reflects the propagation of
photons from one atom to another. The terms  $\propto~1/\zeta_{ab}$
account for retardation effects in the  far-field regime and those  $\propto 1/\zeta_{ab}^3$ account for
instantaneous propagation in the near-field. When $\zeta_{ab}\ll 1$, the elastic $g$ interactions with a strong angular variation are dominant  except close to
the magic angle $\theta_m=\text{arccos}(1/\sqrt{3})$, at which they are  greatly suppressed. In contrast, $ f({\bf r}_{ab})$ is   almost isotropic  in the near-field regime [see Fig.~\ref{fig:fig1}(b)].

 The spatially uniform behavior of  $ f({\bf r}_{ab})$  at short distance is what gives rise to cooperative effects and  superradiant emission \cite{haroche1982}. Under generic conditions, however, superradiance is  a  transient effect that substantially limits the lifetime of dipole excitations. To compensate for the fast decay here we add  an incoherent repumping driving term at a rate $W$. This term is needed to generate a synchronized steady state where long-lasting coherence persists.
  An incoherent repumping drive   is commonly used in laser systems  to maintain population inversion. It can be implemented by coherently driving, at a rate  $\Omega_{\rm ex}$,  the $\ket{\downarrow}$ state to an excited level  that   spontaneously decays, at a rate  $\gamma\gg \Omega_{\rm ex}$, to the state $\ket{\uparrow}$. Due to the fast depletion of the excited state, it can be adiabatically eliminated and thus the  net process  is just an incoherent transfer of   population  from $\ket{\downarrow}$ to  $\ket{\uparrow}$ at a rate $W=\Omega_{\rm ex}^2/\gamma$~\cite{scullybook}.

\begin{figure*}
\centering
\includegraphics[width=1.5\columnwidth]{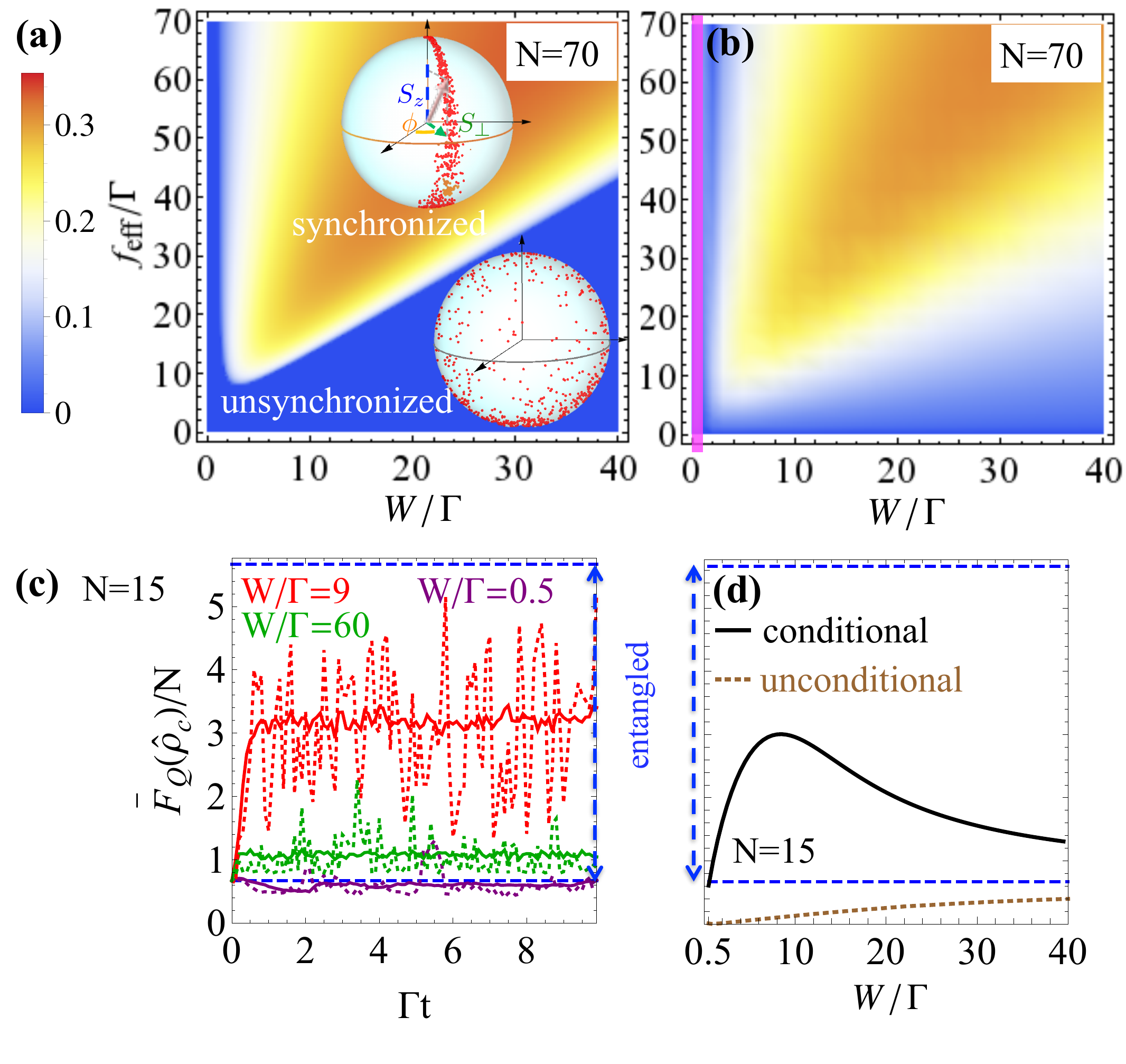}
\caption{(a) Mean-field phase diagram calculated from the order
  parameter $0\leq Z\leq 1/\sqrt{8}$. The insets show snapshots of the tips
  of the Bloch vectors (red points) for dipoles prepared with random
  initial phases and then evolved to steady-state in both regimes. (b)
  Quantum phase diagram calculated from $0\leq Z_Q\leq 1/\sqrt{8}$.
  (c) The time evolution of the conditional  QFI exhibits entanglement (dashed  line: single trajectory,  solid line: mean value of a few trajectories).  Panel (d) shows
  the steady  state QFI vs $W/\Gamma$ after averaging over many trajectories. The solid line corresponds to the conditional case, and indicates entanglement over the repumping range where synchronization exists. Upon computing the ensemble average one recovers the reduced density matrix which leads to a calculated QFI  below the entanglement witness threshold (dashed line). (c) and (d) are shown for $f_{\rm eff}=15 \Gamma$. For
  all panels, $\delta_a=g({\bf r}_{ab})=0$ and $f({\bf r}_{ab})=f_{\rm
    eff}/N$. }\label{fig:fig2}
\end{figure*}

The evolution of  $N$ dipoles is modeled  by a quantum master
equation for the reduced density matrix $\hat{\rho}$ of the dipoles \cite{haroche1982}:
\begin{align}
  \frac{d\hat{\rho}}{dt}&=-\frac{i}{\hbar}[\hat{H}_0,\hat{\rho}]
  +{\mathcal{L}}_{f}[\hat{\rho}]
  +{\mathcal{L}}_W
  [\hat{\rho}],\label{eq:mastermain}\\
 \hat{H}_0&=\frac{\hbar}{2}\sum_{a=1}^N \Big[\delta_a \hat{\sigma}_a^z
  +\sum_{b=1,b\neq a}^Ng({\bf r}_{ab})\hat{\sigma}_a^+\hat{\sigma}_b^-
  \Big],\label{eq:hamil}\\
  \mathcal{L}_{f}[\hat\rho]&=\frac{1}{2}\sum_{a,b}
  f({\bf r}_{ab})(2\hat\sigma_b^-\hat\rho\hat\sigma_a^+
  -\hat{\sigma}_a^+\hat{\sigma}_b^-\hat\rho-\hat\rho\hat{\sigma}_a^+
  \hat{\sigma}_b^-),\!\\
\mathcal{L}_W[\hat\rho]&=\frac{W}{2}\sum_{a}(2\hat\sigma_a^+
\hat\rho\hat\sigma_a^--\hat\sigma_a^-\hat\sigma_a^+\hat\rho-
\hat\rho\hat\sigma_a^-\hat\sigma_a^+).
\end{align}
The Hamiltonian $\hat{H}_0$  generates the coherent evolution of the dipole array where $\hat{\sigma}_a^{(+,-,z)}$ are the Pauli spin operators for
dipole $a$, $\delta_a$ denotes its bare oscillation frequency and
$\hbar$ is the reduced Planck constant. The Lindblad operator functionals, ${\mathcal{L}}_{f,W}$,  describe the inelastic photon emission and incoherent repumping processes, respectively.

\section{Mean-field treatment  and connection to  the Kuramoto model}
To obtain a qualitative picture of how synchronization can happen among the dipoles, we first perform a mean-field  treatment and show the close connection between our quantum model and the prototype models for classical synchronization. The mean-field approach assumes uncorrelated dipoles, i.e., $\hat{\rho}=\bigotimes_a {\hat \rho}_a$, where each
${\hat\rho}_a=\sum_{\sigma,\sigma'=\uparrow,\downarrow}
\rho_a^{\sigma,\sigma'}|\sigma\rangle\langle\sigma'|$ is a $2\times2$
matrix in the pseudospin $1/2$ basis $\{\ket{\uparrow},
\ket{\downarrow} \}$.  The components of the single-dipole density
matrix, $\hat{\rho}_a$, can be visualized as a Bloch vector $\{S_a^{\perp}(t)\cos{\phi_a(t)},S_a^{\perp}(t)\sin{\phi_a(t)},S_a^z(t) \}=
(1/2)\{\rho_a^{\uparrow\downarrow}+ \rho_a^{\downarrow\uparrow},-i(\rho_a^{\uparrow\downarrow}- \rho_a^{\downarrow\uparrow}), \rho_a^{\uparrow\uparrow}-\rho_a^{\downarrow\downarrow}\}$
(Fig.~\ref{fig:fig2}). The mean-field solution yields a system of coupled
nonlinear differential equations for $\rho_a^{\sigma,\sigma'}$. For each $a=1,2,\dots, N$  the parameters evolve as
 \begin{align}
&\frac{dS^z_a(t)}{dt}=-S^\perp_a(t)\sum_{b\neq a}S^\perp_b(t)
\Big[f({\bf r}_{ab})\text{cos}[\delta\phi_{ba}]\nonumber -g({\bf r}_{ab})
\text{sin}[\delta\phi_{ba}]\Big] \\&\quad
-\Gamma
\left[\frac{1}{2}+S^z_a(t)\right]+W\left[\frac{1}{2}-S^z_a(t)
\right],\label{eq:mfsync1}\\
&\frac{dS^\perp_a(t)}{dt}=- S^z_a(t)\sum_{b\neq a}S^\perp_b(t)
\Big[g({\bf r}_{ab})\text{sin}[\delta\phi_{ba}]\!-f({\bf r}_{ab})
\text{cos}[\delta\phi_{ba}]\Big] \nonumber \\ &\quad-\!\frac{\Gamma+W}{2}S^\perp_a(t) ,\\
&\frac{d\phi_a}{dt}=\delta_a\!+\!\!\!\!\sum_{b=1,b\neq a}^N\!\!\!\!S_a^{z}\frac{S_b^{\perp}}{S_a^{\perp}}
  \Big[g({\bf r}_{ab})\text{cos}[\delta\phi_{ba}]+\!f({\bf r}_{ab})\text{sin}[\delta\phi_{ba}]\Big], \label{eq:mfsync3}
\end{align}
where $\delta\phi_{ba}(t)=\phi_b(t)-\phi_a(t)$. The term proportional to $f({\bf r}_{ab})$ in Eq. (\ref{eq:mfsync3}) that contains the sine
function can be identified with a similar term in the Kuramoto
model (KM)~\cite{kuramoto}:
 \begin{eqnarray}
\frac{d\phi_a}{dt}=\delta_a+K\sum_{b=1}^N\text{sin}[\delta\phi_{ba}], \end{eqnarray}
where $K$, the coupling strength per oscillator, must be large enough and positive for
synchronization to occur.  The term proportional to $g({\bf r}_{ab})$
that contains the cosine function  appears in the Sakaguchi-Kuramoto model~\cite{Sakaguchi}---a more general but similar synchronization model to the KM. Compared to the basic KM, the situation here is more
complex. This is due to the fact that in Eq.~(\ref{eq:mfsync3}) the
coupling constants are nonuniform and effectively time-dependent,
since $S_a^\perp(t)$ and $S_a^z(t)$ are dynamic variables.

 To investigate whether the mean-field model admits
spontaneous synchronization we consider first the simplified case
where $\delta_a=0$ for all dipoles, impose $g({\bf r}_{ab})=0$ for all
pairs, and assume a constant collective decay rate $N f({\bf
  r}_{ab})\equiv f_{\rm eff}$. We define a   global order parameter $Z$  as
$Ze^{i\Phi}=\frac{1}{N}\sum_aS^\perp_a e^{i\phi_a}$ and look for a solution in which $Z$ is time-independent and synchronized oscillators possess a
collective frequency $\overline{\omega}$, and thus a macroscopic phase
$\Phi=\overline{\omega} t$.   These conditions lead to two
equations for the order parameter $Z$ and the collective frequency
$\overline{\omega}$ (see Appendix \ref{appendixa}):
\begin{align}\label{zetas}
{\overline{\omega}}&=0,\\
Z &= \frac{\sqrt{f_{\rm eff}(W-\Gamma)-(W+\Gamma)^2}}{\sqrt{2}f_{\rm eff}}.
\end{align}

 The solution is shown in Fig.~\ref{fig:fig2}(a). The insets show the phase
distribution in the steady-state for an array of oscillators initially
prepared with a random distribution of phases for two different values
of the repump rate $W$. For a slow repumping rate (bottom inset), the system remains
unsynchronized. As the repumping rate is increased beyond a threshold
value, the system enters a synchronized state, as can be seen by the
appearance of phase locking and the resulting narrow phase
spread (top inset).  This can be explained by the fact that one necessary condition
for synchronization in the KM is $K>0$, which translates to the
requirement $S_a^z>0$ on average in our model and thus the need to
have sufficiently large repump rate. In the limit $f_{\rm eff} \gg \Gamma$ (e.g., for large N), maximum  synchronization is achieved at $W_{\rm opt}=f_{\rm eff}/2$, where  the  order parameter $Z$ reaches a  maximum   value
$Z_{\text{max}}\approx\sqrt{1/8}$. For this optimal
condition  for synchronization the quantum dipoles are ordered with the same phase and   radiate with atomic inversion $S_a^z\approx1/4$.  Note that the maximum order parameter is smaller than  $1/2$ even when fully synchronized because of this required finite value of the atomic inversion. One intriguing aspect is that repumping, which is the process that builds up synchronization, is  itself an incoherent process. It is crucial that repumping does not preserve the norm of the collective Bloch vector, allowing it to extend or contract. For large $W>W_{\text{opt}}$, $Z$ decreases again reflecting a suppression of synchronization. In this limit the dipoles are repumped so fast  that they are  all driven to the $\ket{\uparrow}$ state ($S_{a}^z\to 1/2$ and $S_{a}^{\perp}\to 0$) and  phase coherence between them  cannot build up.

 The cases of a heterogeneous distribution of $\delta_a$'s or finite $g({\bf r}_{ab})\neq0$ can be treated at the mean-field level in a simple way. The results,  summarized in the Appendix \ref{appendixa},  are qualitatively similar. In this case we define the effective couplings as $N f({\bf
  r}_{ab})\equiv f_{\rm eff}$ and
$N g({\bf
  r}_{ab})\equiv g_{\rm eff}$. In general, the inclusion of a finite spread $\Delta$ in $\delta_a$ decreases the value of the order parameter $Z$. For instance,
if $\delta_a$ is sampled from a Lorentzian distribution
$p(\delta_a)=\Delta/[\pi(\Delta^2+\delta_a^2)]$,
\begin{equation}
Z=\frac{\sqrt{f_{\text{eff}}P-Q^2
+2\Delta^2-2\Delta\sqrt{\Delta^2+f_{\text{eff}}P}}}
{\sqrt{2}f_{\text{eff}}},
\end{equation} where $Q=\Gamma+W$ and $P=W-\Gamma$.
Optimal synchronization is obtained at a smaller repumping rate,
$W_{\text{opt}}\approx f_{\text{eff}}/2-\Delta/\sqrt{2}$. We note that
for given $f_{\text{eff}}$ and $W$, synchronization is destroyed (that
is, $Z=0$) at
$\Delta_c=(
Q^2-f_{\text{eff}}P)/(2Q)$.

 When the dipoles have identical detunnings, $\delta_a=0$, the elastic couplings  simply induce a global frequency shift ${\overline{\omega}}=g_{\text{eff}}Q/(2f_{\text{eff}})$
that can be eliminated by moving to a rotating frame.

\section{Quantum synchronization for the collective system}
In the simplified case  where $\delta_a=0$ for all dipoles,  $g({\bf r}_{ab})=0$ for all
pairs, and  a constant collective decay rate $N f({\bf r}_{ab})\equiv f_{\rm eff}$,  it is possible to exactly solve
Eq.~\eqref{eq:mastermain}, {\it i.e.}, the full quantum dynamics, even for many
particles, allowing us to benchmark the validity of the mean-field
solution. This is due to the invariance of the master equation under
individual dipole permutations that reduces the scaling of the
Liouville space from exponential, $4^N$, to polynomial, of order
$N^3$~\cite{su4}.

\subsection{Phase Diagram}
  Quantum fluctuations can lead to phase diffusion
and to decay of single particle coherences in the steady-state (it is
possible for $\langle {\hat \sigma}_a^+\rangle \to 0$ even in a
synchronized state), so $Z$ cannot be used as a measure of synchronization in a beyond mean-field
treatment. However, phase locking in quantum mechanics can be
quantified by the degree of spin-spin correlations $Z_Q$, defined by
$Z_Q^2\equiv \langle\overline{{\hat \sigma}_a^+{\hat
    \sigma}_b^-}\rangle$, where the bar indicates an average over all
pairs of different dipoles $a$ and $b$.  For an unsynchronized state
$Z_Q$ is $0$ and for a completely synchronized state $Z_Q$ is
$Z_Q^{\rm max}=1/\sqrt{8}$ ~\cite{dmeiser2,suplaser}. The
corresponding phase diagram, shown in Fig.~\ref{fig:fig2}(b), closely
resembles the mean-field one.

To demonstrate that  $Z_Q$ can be used to quantify the  emergence of quantum synchronization, regardless of the inherent non-equilibrium and dissipative  character of our system, we  have also computed pairwise two-time correlation
functions (see Appendix B). The decay rate  of these correlations encodes information about the spectral coherence of the emitted radiation. The range of $W/\Gamma$ values where the emitted light is maximally coherent agrees with the regime where the system is optimally synchronized according to $Z_Q$. Moreover, we have also confirmed  the moderate importance of higher order correlations in the synchronized steady-state by comparing the exact solution with a  cumulant expansion calculation (which includes lowest order corrections to the mean-field result). We find the cumulant expansion agrees well with the exact solution (see Appendix B). The only limit  where there are important deviations is at very weak pumping $W\ll \Gamma$ where the system favors subradiant emission  arising from strong atom-atom correlations [indicated by the purple region in Fig.~\ref{fig:fig2}(b)]~\cite{su4}.

\subsection{Quantum correlations and entanglement}

\begin{figure*}
\centering
\includegraphics[width=1.7\columnwidth]{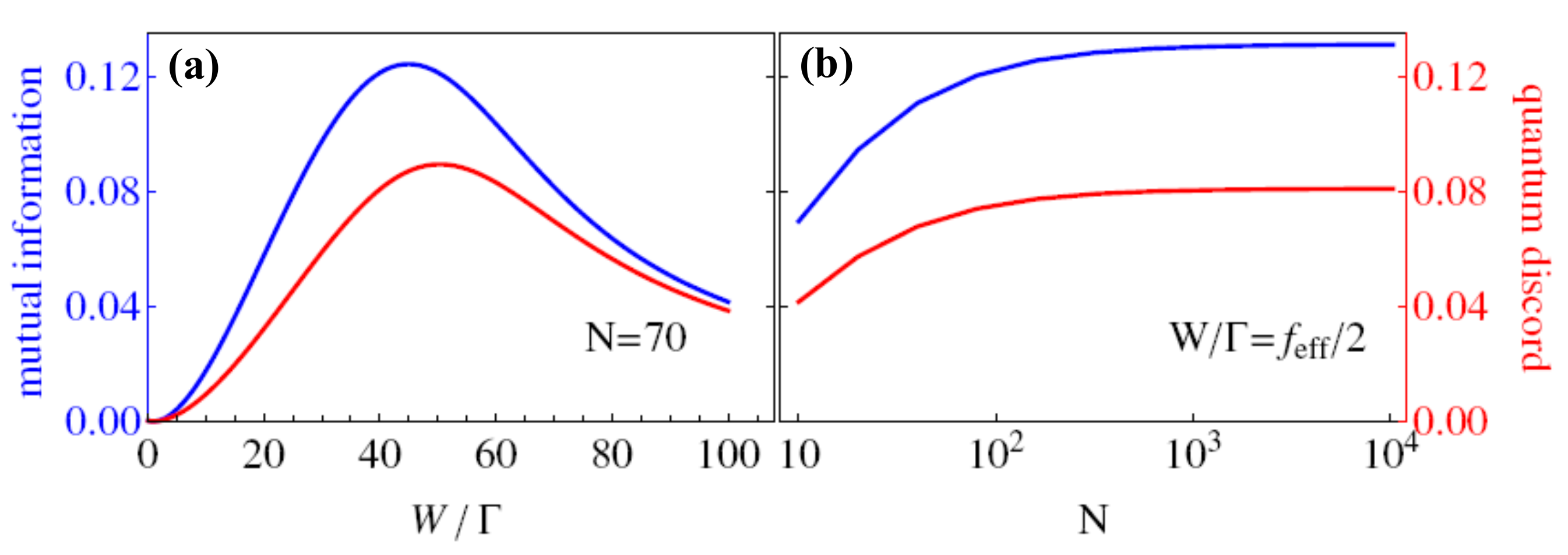}
\caption{
  Quantum correlations and total correlations. The total and quantum correlations in the steady state are quantified by the mutual information $0\leq \mathcal{I}\leq 2 $ and quantum discord  $0\leq \mathcal{D}\leq 2$.  (a) In the synchronized phase, there are nonzero quantum correlations and classical correlations ($\mathcal{I}-\mathcal{D}$), and both show a dependence on $W$ that qualitatively agree with $Z_Q$. (b) Even in the thermodynamic limit, quantum correlations remain a significant fraction of the total correlations. For both panels, $\delta_a=g({\bf r}_{ab})=0$ and $f({\bf r}_{ab})=f_{\rm
    eff}/N$. }\label{fig:fig2s}
\end{figure*}
The robust macroscopic quantum coherence exhibited by the synchronized
 state leads to the natural  question of whether or not entanglement can be
present in the steady-state even in this dissipative environment. Most previous studies that attempted to address this question have been limited to  small systems~\cite{Giorgi2013,Lee2013,Manzano2013,optoprl2013} and  focused on the entanglement between a pair of synchronized oscillators. Here, to determine the non-separability of the many-body steady-state, we compute the average of the quantum Fisher information (QFI) and use it
as an entanglement witness~\cite{Strobel2014,Sup,fishersmerzi}.   Any $N$ particle
state with $(N^2+2N)/3\geq \bar{F}_Q(\hat{\rho})>2N/3$ is entangled
(non-separable) and a quantum resource for phase estimation (see Appendix C for details).

Due to dissipation, the density matrix of the system is reduced to a mixed state, as obtained from Eq.~(\ref{eq:mastermain}), which describes the dynamics of the system after a statistical average over many experimental trials.  However, the evolution of the system for an individual experimental realization can be quite different. We consider a {\em Gedanken}\/ experiment in which one monitors the system evolution and keeps  a measurement record of the emitted photons. The  evolution of the system is then conditioned on the measurement record~\cite{carmichael,qsdreview}. This type of conditional evolution  has been widely studied in quantum optics and utilized together with quantum feedback control in examples such as the optimal generation of spin squeezed  states (See Refs.~\cite{feedbacksqueezing,feedbackreview}). It should be emphasized that the conditional evolution based on the measurement record gives a quantum trajectory that should not be regarded simply as a numerical tool to allow the efficient assembly of ensemble averages. Each quantum trajectory is a potentially realizable physical outcome (even if hard to perform in practice) as allowed by the quantum dynamics of the open quantum system under consideration.

We calculate $\bar{F}_Q(\hat{\rho}_c)$ (with the $c$ in
$\hat{\rho}_c$ meaning conditional) for each conditional trajectory and in Fig.~\ref{fig:fig2}(d) we show its average over a sufficiently large set of trajectories at steady state (see Appendix C). For this conditional evolution we observe entanglement in a parameter regime that correlates with $Z_Q>0$ [see Fig.~\ref{fig:fig2}(c) and (d)].  On the other hand,  if we discard the information
present in the measurement record, by using the ensemble averaged $\rho$ obtained from directly solving Eq.~(\ref{eq:mastermain}), and then computing $F_Q(\hat{\rho})$, the QFI falls below the entanglement witness threshold [see Fig.~\ref{fig:fig2}(d)].

To differentiate quantum effects from classical ones, we further calculate the quantum discord $\mathcal{D}$~\cite{Sup}, which can be considered as a measure of quantum correlation more general than entanglement and more robust in a dissipative environment~\cite{discordzurek,discordNatphys2012,discord2001}. Separable states with nonzero $\mathcal{D}$ are intrinsically nonclassical, since local measurements performed on a subsystem inevitably disturb the whole system~\cite{discordzurek,Modi2012}.  We  measure classical correlations of the steady state by the difference between the mutual information $\mathcal{I}$~\cite{Sup} and  $\mathcal{D}$. We find the mixed steady-state contains nonzero quantum correlations in the
synchronized regime  Fig.~\ref{fig:fig2s}(a). Moreover we observed  that although  both  $\mathcal{D}$ and $Z_Q$ exhibit a similar dependence with pumping rate $W$, they do not  exactly peak at the same  value ~ \cite{Ameri2014}.

 Although the existence of a nonzero $\mathcal{D}$ has been reported  to exist in several quantum  synchronization studies ~\cite{Manzano2013,Modi2012}, we want to emphasize that those have been always limited to small systems. To our knowledge our calculations are the first to consider  $\mathcal{D}$  in macroscopic samples. In Fig.~\ref{fig:fig2s}(b) we show the dependence of    $\mathcal{D}$  with system size.  Our calculation shows that quantum correlations remain a significant fraction of $\mathcal{I}$ even in the thermodynamic limit.

\begin{figure*}[tb]
\centering
\includegraphics[width=1.8\columnwidth]{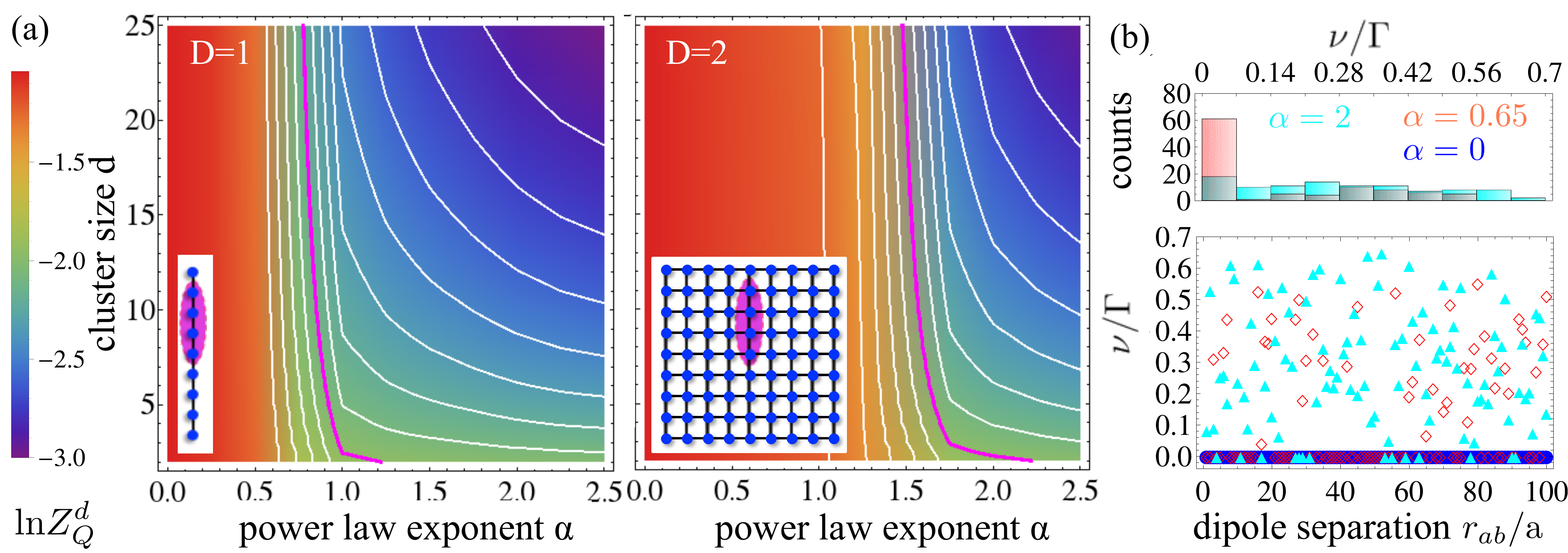}
  \caption{ (a) Spin-spin correlations, $Z_Q^d$, in linear clusters containing $d$ dipoles
    at optimal repumping $W$ for power-law couplings $f({\bf
      r}_{ab})=\frac{\Gamma}{4} \left( \frac{{\rm a}}{{r}_{ab}}
    \right)^{\alpha}$ with lattice spacing ${\rm a}$. We set
    $\delta_a=g({\bf r}_{ab})=0$ and consider  $N=900$ dipoles
    arranged in both  linear ($D=1$) and  square lattice ($D=2$)
    geometries. For $\alpha\lesssim D$ global  synchronization is observed and the order parameter is independent on cluster size $d$. For  $D\lesssim \alpha$  the order parameter starts to clearly decay with increasing $d$. The magenta line ($Z_Q^d=0.14)$   provides
    an indicative scale of the boundary between global and local synchronization.
   The white contour lines  provide an indication of the decrease  of the synchronized domains  with increasing $\alpha$.  (b) Pair wise two-time correlation functions
    in the steady-state are parametrized by $Z_{a,b}(\tau)
    =A\cos(\nu\tau)\exp(-\gamma\tau)$ where $a$ is chosen as the
    central dipole of a linear chain of $N=200$ dipoles. The dipoles are assigned random
    detunings $\delta_a$ distributed uniformly in
    $[-\Gamma/2,\Gamma/2]$. The dark blue, red, and light blue  symbols correspond to $\alpha=0, 0.65$ and 2 respectively. The histogram of frequencies $\nu$ exhibits similar synchronization
    regimes than those seen in (a).   }\label{fig:fig3}
\end{figure*}

\section{Synchronization with finite-range interactions\label{sec:power}}

Up to this point  we have  only considered all-to-all interactions; now we consider the effect of finite range  interactions on synchronization. In the dipole array  both $f({\bf r}_{ab})$  and $g({\bf r}_{ab})$ are  nontrivial functions of ${\bf r}_{ab}$ and contain  terms decaying as a power-law with distance,  $\propto 1/|{\bf r}_{ab}|^{\alpha}$ with $\alpha=1,2,3$. Instead of dealing with  all these terms together, to gain insight on how
spatial inhomogeneities affect quantum synchronization,  we first study   a simpler case assuming a  power-law
cooperative decay $f({\bf r}_{ab})\propto |{\bf r}_{ab}|^{-\alpha}$,
with the exponent $\alpha$ as a variable parameter and set
both  $g({\bf r}_{ab})=0$ and $\delta_a=0$.

 In the classical regime   ~\cite{daido,strogatz1988,powerlaw,uchida},  analytical calculations and numerical simulations considering  arrays of oscillators interacting via  power law interactions on a one-dimensional lattice had  identified  $\alpha_c=3/2$  as the critical value
of the power law exponent  below which long-range phase order is possible~\cite{powerlaw}. For $\alpha < \alpha_c$, a transition to a state in which a finite fraction of the oscillators is entrained takes  place for a sufficiently strong but finite coupling strength in the large system limit. Generalizations of these results to oscillators in $D$ dimensions~\cite{powerlaw} have also  identified three different regimes for synchronization: perfect phase ordering for $\alpha\leq D$, entrainment with long-range phase order
for $\alpha <3D/2$ and a crossover to exponential decay of
correlations at $\alpha= (3D+ 1)/2$. Reference  ~\cite{uchida} has also  suggested that in the regime  $\alpha> D$  global synchronization is absent  but local synchronization
persists for arbitrary weak coupling with a slowly decaying  order parameter.

 To quantify the effect of finite-range interactions on synchronization in the quantum regime we compute spin-spin correlations within linear clusters that contain $d$ dipoles, $ (Z_Q^d)^2\equiv \langle\overline{{\hat \sigma}_a^+{\hat
    \sigma}_b^-}\rangle_d$, using a cumulant expansion method as described in Appendix~B. Here the bar followed by a subscript $d$ indicates an average over the
pairs of different dipoles $a$ and $b$ contained in a linear cluster of size $d$. The linear clusters start at the central spin as shown in
Fig.~\ref{fig:fig3}. We have confirmed  that the cumulant expansion method reproduces well the correlation functions by performing direct comparisons with the exact solution (see Appendix B). Fig.~\ref{fig:fig3}(a) shows the behavior of $ Z_Q^d$ as a
function of cluster size $d$ and power-law decay exponent  $\alpha$ in arrays of dimension $D=1$ and 2.
Clear global synchronization  with an order parameter $Z_Q^d$ independent of $d$ is observed for $\alpha \lesssim D$.
For $0<\alpha\lesssim D/2$, the local order parameter $ Z_Q^d$ is almost independent of $\alpha $ and $d$ and the system behaves almost like the all-to-all system.  For $D/2\lesssim\alpha\lesssim D$ synchronization remains global and  almost independent of $d$, but the order parameter  slowly decreases with $\alpha$. For $\alpha\gtrsim  D$, synchronization becomes local and correlations quickly decrease with cluster size.  The magenta contour provides an indicative scale of the boundary between global and local synchronization. The white contour lines also provide information about the decrease of  the  order parameter with increasing $\alpha$ and $d$. We observe that, as in the classical case, $\alpha \sim D$ roughly marks the transition between global and local synchronization, although a more quantitative comparison would require far larger systems.

An alternative way to characterize domain formation and the fact that
it can  persist even when there is a variation in the local
detunings, $\delta_a\neq 0$, is to examine pairwise two-time correlation
functions, $Z_{a,b}(\tau) \equiv\lim_{t\rightarrow\infty}\langle (\hat
\sigma_a^+ (t+\tau) + \hat\sigma_b^+(t+\tau))(\hat \sigma_a^-(t) +
\hat\sigma_b^-(t))\rangle$, which can be related to the emission spectrum of the pair of atoms~\cite{carmichael}. The oscillations in  $Z_{a,b}(\tau)$ encode information about the relative precession rate between different dipoles. By parameterizing $Z_{a,b}(\tau)$ as $Z_{a,b}(\tau)
    =A\cos(\nu_{ab}\tau)\exp(-\gamma\tau)$ we can extract the relative precession frequency $\nu_{ab}$ between dipoles  $a$ and $b$, where entrainment of dipoles $a$ and $b$ corresponds to $\nu_{ab} = 0$. To explore the entrainment of dipole pairs in our system, we assign random detunings distributed uniformly in $[-\Gamma/2,\Gamma/2]$ to a linear chain of $N = 200$ dipoles and calculate $\nu_{ab}$ for $b = 1,2,\dots,100$ with $a = 101$ corresponding to the central dipole in the chain.  Synchronization regimes similar to those shown in Fig.~\ref{fig:fig3}(a) are observed for this $D=1$ system, which we illustrate in Fig.~\ref{fig:fig3}(b) by plotting a histogram (top panel) and the distribution of frequencies $\nu$ (bottom panel) for three values of $\alpha$. For global coupling, $\alpha = 0$ (dark blue symbols), all the dipoles become entrained with each other ($\nu = 0$), indicating complete synchronization; for $\alpha = 0.65$ (red symbols) dipoles split into entrained ($\nu = 0$) and drifting ($\nu \neq 0$) groups. While not all dipoles are entrained, the entrained dipoles are distributed along the whole array, and thus synchronization is still global; and for $\alpha = 2$ (light blue symbols) the majority of dipoles are not entrained. These observations of relative precession frequencies between pairs of oscillators are consistent with the regimes obtained from the order parameter plotted in Fig.~\ref{fig:fig3}(a).

\section{Synchronization of dipoles with elastic interactions}
\begin{figure}[tb]
\centering
\includegraphics[width=\columnwidth]{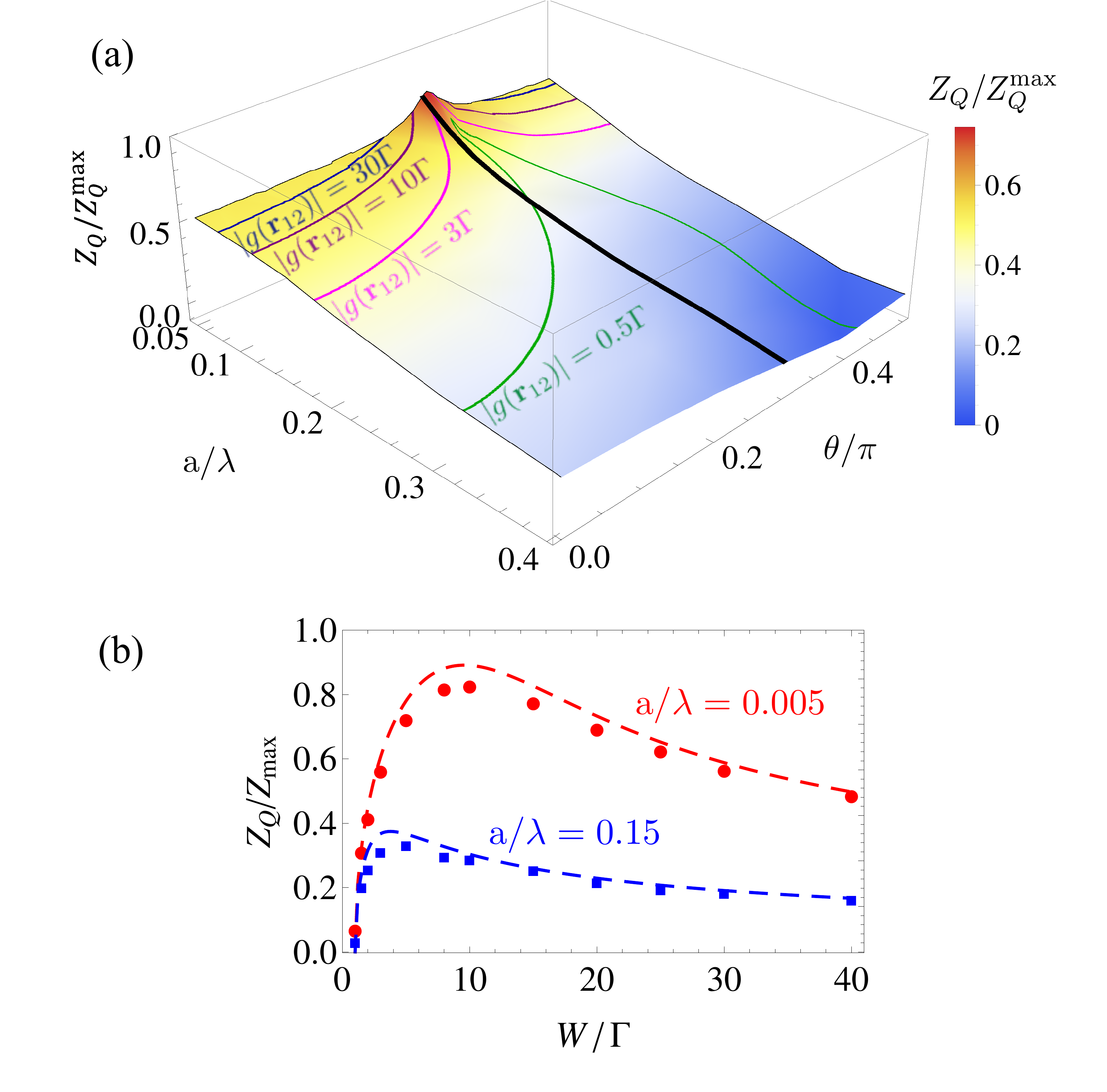}
  \caption{(a) Synchronization in dipole arrays is demonstrated for $N=12$ dipoles on a
    line when subjected to incoherent pumping (optimal rate). In this geometry,  regardless of the strong angular variation of $g$  with the lattice spacing $\rm{a}$ (see contours)   the  order parameter, $Z_Q$,
    (normalized by $Z_Q^{\rm max}=1/\sqrt{8}$) exhibits a weak dependence on $\theta$ and $\rm{a}$ and reaches a maximum at $\theta=\theta_m$. (b) The order parameter is computed for $N=16$ dipoles on a line with $\theta=\theta_m$ and $f_{\rm eff}=\sum_{a,b\neq a}f({\bf r}_{ab})/(N-1)$ (symbols), and for a system with constant $f({\bf r}_{ab})=f_{\rm eff}/N$ and $g({\bf r}_{ab})=0$ (dashed line)s. Similar dependence on $W$ is found for these two different systems. Here the order parameter for dipoles is always smaller in the presence of elastic interactions.
  }\label{fig:fig4}
\end{figure}

We now treat the full problem of radiating quantum dipoles
incorporating elastic interactions $ g({\bf r}_{ab})$ and the intricate competition of
spatially-dependent  and anisotropic couplings [both $ g({\bf r}_{ab})$ and $ f({\bf r}_{ab})$ have terms with  power law dependence $\alpha=1,2,3$ on distance] (Fig.~\ref{fig:fig1}). We  solve the full master equation without any approximation \cite{carmichael} for systems of up to twenty dipoles in a chain using the actual spatial dependence of both $ f({\bf r}_{ab})$ and $ g({\bf
  r}_{ab})$, and set $\delta_a=0$. We observe a robust synchronized state that exists in a wide parameter
space.  As long as $f_{\rm eff}\equiv N \overline{f({\bf
  r}_{ab})}$ is large enough, we find that
synchronization takes place and is only weakly affected by substantial
differences in $ g({\bf r}_{ab})$, {e.g.}, variations  in the dipole array that modify $g({\bf r}_{ab})$ by   two orders of magnitude  only decrease the
order parameter by a factor of two (Fig.~\ref{fig:fig4}) in the steady state. This is in striking contrast to the situation in a system without dissipation, where the elastic interaction is  known to generate entanglement between spins  and to cause a decay of the order parameter during time evolution~\cite{spinsqueezing}.

For the orientation $\theta_m=\arccos(1/\sqrt{3})$, the order parameter reaches a significant fraction of $Z_Q^{\rm max}$, indicating the emergence of macroscopic  spontaneous synchronization of
the radiating quantum dipole array (Fig.~\ref{fig:fig4}).
 To further  emphasize the relevant role played by the inelastic term, in Fig.~\ref{fig:fig4}(b) we compare a solution of  the master equation [Eq.~(\ref{eq:mastermain})]  for two cases: a system of  coupled dipoles  arranged  in a 1D chain and oriented at the magic angle (symbols) and an array of identically coupled dipoles with the same  $f_{\rm eff}$ but experiencing only inelastic interactions [$g({\bf r}_{ab})=0$, dashed lines]. The calculated order parameters agree  well for the two different cases. The similar behavior  demonstrates  that in spite of the complex geometry of the dipolar interactions, the capability of the dipole system to synchronize can be characterized to great extent by the quantity $f_{\rm eff}$.

\section{Experimental implementation}
Our calculations above demonstrate the potential for synchronization in a dense array of dipoles. The flexible and precise control exhibited by ultracold atomic systems make them ideal  platforms to experimentally investigate the synchronization phenomenon predicted here. Atomic systems operate with  a large number of quantum oscillators and also allow for the tunability  of the  interaction parameters over a broad range.

One possible set-up   to observe synchronization consists of arrays of ultracold ${}^{87}$Sr atoms prepared in two electronic internal states that form the two-level system. The $\ket{\downarrow}$ could then correspond to the long-lived  $5s5p~{}^3\!P_0$ state, with an intercombination line narrower than $10^{-3}$ ${\rm{s}^{-1}}$ . This is the state used to operate  the most precise  atomic clocks~\cite{Bloom2014}. The $\ket{\uparrow}$ could correspond to the $5s4d~{}^3\!D_1$ state with a natural linewith $\Gamma=290\times 10^3$ ${\rm{s}^{-1}}$. Both states can be trapped in an optical lattice at the magic wavelength ${\rm a}=0.2$ $\mu$m~\cite{Olmos2013}, that generates the same trapping potential for both states minimizing Stark shifts and inhomogeneities in the coupling constants. The dipole-dipole interactions are mediated by photons at the  wavelength $\lambda=2.6$ $\mu$m and thus,   as shown in Fig.~\ref{fig:fig4}, the ratio $a/\lambda<0.08$ falls in the parameter regime where dipoles can be synchronized.

By changing the angle between  the laser beams used to form the lattice potential, the lattice spacing can be varied  allowing tunability of the interaction strength between dipoles. The incoherent pumping can be realized by coherently transferring the $5s5p~{}^3\!P_0$ population to one or several appropriate intermediate states that decay rapidly to the $5s4d~{}^3\!D_1$ state~\cite{dmeiser1}. An example would be the $4d5p~{}^3\!P_1$ state~\cite{srdata}.

The polarization of the dipoles can be oriented in an arbitrary direction by an electromagnetic field. Although  all the dipoles cannot be oriented at the magic angle in a 3D geometry, one  may still suppress the elastic interactions by dynamical decoupling techniques adopted from NMR \cite{NMR}. Those have been already demonstrated in ultracold polar molecule  systems \cite{yan2013}. Another possibility is to use
 a spatial configuration of external fields that induces an `averaging out' of the dominant elastic interactions~\cite{becgravitysom}. Moreover, by slightly departing from the magic-wavelength condition, the dipoles can be  subjected to  onsite inhomogeneities that generate  different detunings $\delta_a$.

  The phase synchronization can be probed  by measuring $Z_Q$, which experimentally  can be directly obtained from the fluorescence intensity. As suggested in Sec.~\ref{sec:power}, phase locking can also be extracted from two-point correlations which can be determined by analyzing the fluorescence spectrum ~\cite{dmeiser1}.

  An intriguing but also more speculative and less controllable realization of our  quantum dipole model  is the case of fluorescent organic molecules. A possible two-level configuration in those systems consists of a vibrational level of the ground electronic potential chosen as $\ket{\downarrow}$ and the lowest vibronic level of the first excited potential chosen as $\ket{\uparrow}$. Incoherent pumping can be realized by driving an optical transition to a higher excited vibronic level $\ket{\phi}$ in the first excited potential, which decays on picosecond timescales to the state $\ket{\uparrow}$ via non-radiative transitions \cite{Leising:2000}. Typical values of the fluorescence wavelength $\lambda_{\rm f}$ and lifetime $\tau_{\rm f}$ for organic chromophores under a variety of environmental conditions put these systems in a regime of near optimal synchronization. For instance,  pseudoisocyanine chloride (PIC) and merocyanine derivatives  commonly used in organic light-emitting diodes (LED)~\cite{Dorn:1986,Berlepsch:2000,Kato:2005,Aramendia:1988} typically form low-dimensional molecular aggregates in liquid solution with ${\rm a}\approx 0.5 - 2.0\;{\rm \AA}$, and ratios ${\rm {a}}/\lambda_{\rm f}$ on the order of $10^{-3}$. The typical fluoresence decay rate for these organic chromophores is $\Gamma\sim 0.1-1$ GHz \cite{Leising:2000}. In order to achieve $W/\Gamma = 1$ and enter the synchronized phase, the required pumping laser intensity is $I_W\sim 1-10$ kW/cm$^2$, which is lower than the theoretical lasing threshold intensities $I_{\rm th}\sim 0.1-1$ MW/cm$^2$ of dye lasers \cite{Leising:2000}. Therefore, it should be feasible to achieve steady state synchronization of organic dipoles via incoherent optical driving.

\section{Conclusion}
We have demonstrated that a system of radiating quantum dipoles can be synchronized
in the presence of repumping. Our analytic mean-field approach provides a direct analogy between synchronization  of quantum dipoles and synchronization  of  classical phase oscillators. Using exact solutions of the master equation  and a cumulant expansion approach,  we determined  the necessary conditions for   synchronization, and the entanglement properties in the steady state of  macroscopic ensembles under different measurement protocols. We also  analyzed the  effect of finite-range interactions in  large arrays. To our knowledge those have been previously explored only in the classical regime. For treating  the general case of dense packed dipoles, we numerically solved the master equation exactly for up to twenty dipoles, and studied the effect of anisotropic elastic interactions.

Our results show that the intrinsic macroscopic coherence of the superradiant
steady state is inherently resilient to single particle decoherence, spatial inhomogeneities, and
noisy environmental effects. This observation could  have relevant application to the   development of low-threshold organic lasers, highly efficient
solar cells,  materials with enhanced chemical reactivity, as well as ultra-precise quantum devices, where these effects are anticipated to play an important role. Moreover, since quantum  synchronization is  imprinted in the spectral purity of the emitted radiation \cite{dmeiser2}, the generated light  may potentially serve as a  direct diagnostic  tool of quantum coherences in generic systems beyond cold gases such as organic molecules.

\section{  Acknowledgements:} The authors wish to acknowledge useful discussions with David Nesbitt, James K.  Thompson, Alexey V. Gorshkov, Emanuel Knill, Kaden R. A. Hazzard, Michael Foss-Feig, Zhe-Xuan Gong,  Michael L. Wall, Dominic Meiser, Xibo Zhang and Timur V. Tscherbul. This work was supported by NIST, the NSF (PIF-1211914 and PFC-1125844), AFOSR, AFOSR-MURI, ARO individual investigator awards, and DARPA QuASAR. Computations utilized the Janus supercomputer, supported by NSF (award number CNS-0821794), NCAR, and CU Boulder/Denver.

\appendix
\setcounter{figure}{0}
\setcounter{equation}{0}
\renewcommand{\appendixname}{ APPENDIX}
\makeatletter\renewcommand{\fnum@figure}{\figurename~A\thefigure}

\section{{Mean-field approach}}\label{appendixa}

The mean-field ansatz, $\hat{\rho}=\bigotimes_a{\hat \rho}_a$, reduces
the dynamics to $3N$ coupled nonlinear differential equations presented  in the main text.    In the most generic case we define local order parameters to take
into account the effect of the inhomogeneous couplings:
\begin{equation}
  X_ae^{i\Phi}=\sum_{b\neq a}f({\bf r}_{ab})S^\perp_b
e^{i\phi_b},~ Y_ae^{i\Phi}=\sum_{b\neq a}g({\bf r}_{ab})S^\perp_b
e^{i\phi_b}.\nonumber
\end{equation} If  the local order parameters vary slowly over the system size, and can be approximated to be the same for all dipoles one can define  $X_a\approx f_{\text{eff}}Z,~  Y_a\approx g_{\text{eff}}Z$, where the  global order parameter $Z$ is defined as
$Ze^{i\Phi}=\frac{1}{N}\sum_aS^\perp_a e^{i\phi_a}$ and the effective couplings are given  by $f_{\text{eff}}=\sum_a\sum_{b\neq a}f({\bf r}_{ab})/(N-1)$ and
$g_{\text{eff}}=\sum_a\sum_{b\neq a}g({\bf r}_{ab})/(N-1)$.

The steady-state solution $\dot Z = 0$, $\Phi = \bar \omega t$ leads to two self-consistent
equations for the order parameter $Z$ and the collective frequency
$\overline{\omega}$
\begin{align}
&\hspace{-1.7ex} Z =\!\sum_{a}^N\frac{ ZP[f_{\text{eff}}Q+2g_{\text{eff}}(\delta_a+
\overline{\omega})]}{NQ[4(\delta_a+\overline{\omega})^2
+(2f_{\text{eff}}^2Z^2+2g_{\text{eff}}^2Z^2+Q^2]},\\
&\hspace{-1.7ex} 0=\!\sum_a^N\frac{ ZP[g_{\text{eff}}Q-2f_{\text{eff}}(\delta_a+
\overline{\omega})]}{NQ[4(\delta_a+\overline{\omega})^2
+(2f_{\text{eff}}^2Z^2+2g_{\text{eff}}^2Z^2+Q^2]},
\end{align}
which can be evaluated in the $N\to \infty$ limit as integrals when the detunnings $\delta_a$ have a known distribution.

\section{{ Cumulant expansion approach and two-time correlation between dipoles}}
\begin{figure}[t]
\includegraphics[width=0.5\textwidth]{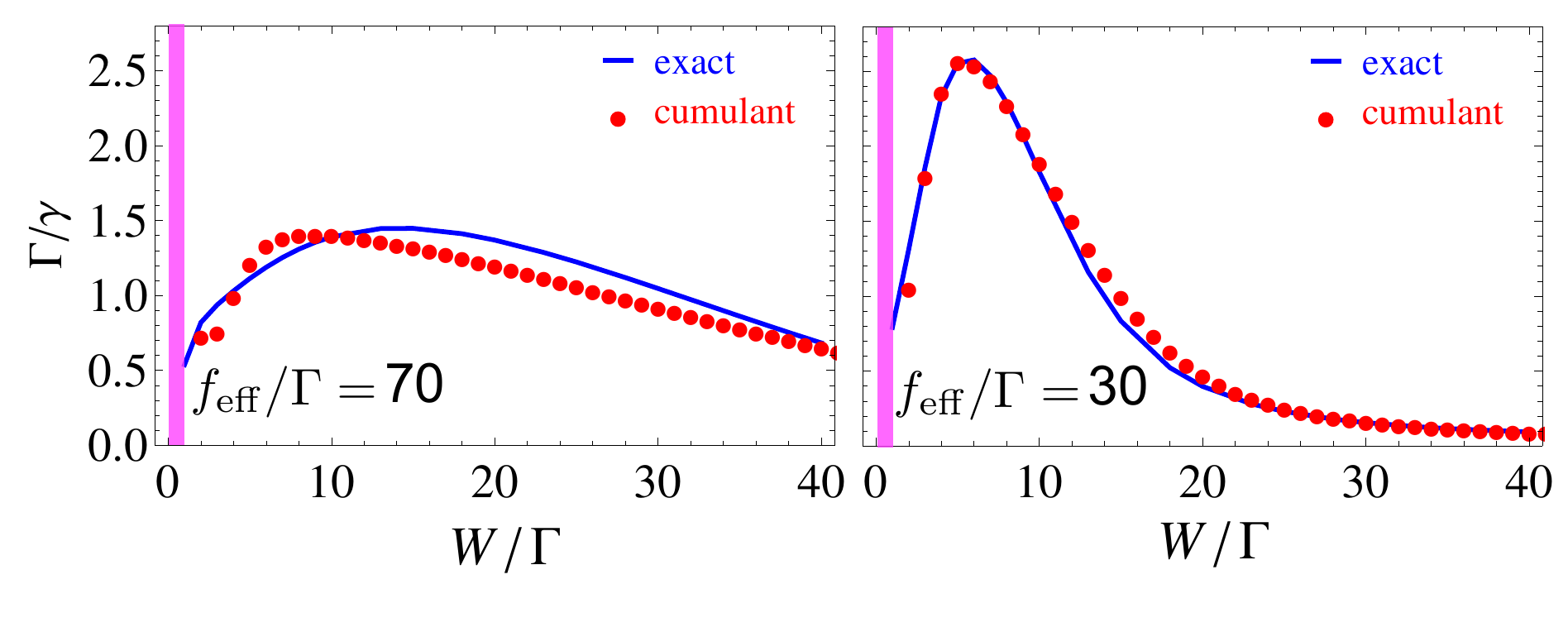}
  \caption{ \textbf{Pair-wise two-time correlation functions in the steady-state parametrized by} $Z_{1,2}(\tau) = A{\rm exp}(-\gamma\tau)$.  The correlations are calculated for a pair of dipoles in an ensemble of $N=70$ dipoles identically coupled with $f({\bf r}_{ab})=f_{\rm eff}/N$.   The cumulant expansion solution agrees with the full calculation except at $W \ll\Gamma$ where subradiant behavior dominates (purple region)~\cite{dmeiser2}.}\label{subfig:twotime}
\end{figure}

The cumulant expansion method is a useful theoretical tool for
including correlation effects beyond the mean-field approximation~\cite{dmeiser2,MXramsey,Helmut}.  We keep two-point correlations such as
$\langle \hat \sigma_a^{+,-,z}\hat \sigma^{+,-,z}_b\rangle$, but factorize three-point correlations and higher~\cite{kubo}:
\begin{eqnarray}
  \langle{\hat \sigma}_a^\alpha{\hat \sigma}^\beta_b
  {\hat \sigma}_c^\gamma\rangle&=&\langle{\hat \sigma}_a^\alpha{\hat
    \sigma}_b^\beta\rangle\langle{\hat \sigma}_c^\gamma\rangle+
  \langle{\hat \sigma}_a^\alpha\rangle\langle{\hat \sigma}_b^\beta{\hat
    \sigma}_c^\gamma\rangle+\langle{\hat \sigma}_a^\alpha{\hat \sigma}_c^\gamma
  \rangle\langle{\hat \sigma}_b^\beta\rangle\nonumber\\&&-2\langle{\hat \sigma}_a^\alpha
  \rangle\langle{\hat \sigma}_b^\beta\rangle\langle{\hat \sigma}_c^\gamma
  \rangle.
\end{eqnarray}
This factorization closes the set of dynamical equations of motion for
all single particle observables $\langle{\hat
  \sigma}_a^{+,-,z}\rangle$ and equal time two-point
correlations.  Two-time correlation
functions can be computed by solving~\cite{carmichael}:
\begin{align}
  \frac{d\langle{\hat \sigma}_a^+(t+\tau){\hat \sigma}^-_b(t)\rangle}
  {d\tau}&=-\Big [i\delta_a+\frac{\Gamma+W}{2}\Big]\langle{\hat \sigma}_a^+
  (t+\tau){\hat \sigma}_b^-(t)\rangle\nonumber\\&+\frac{1}{2}f_{ab}
  \langle{\hat \sigma}_a^z(t)\rangle\langle{\hat \sigma}_b^+
  (t+\tau){\hat \sigma}_b^-(t)\rangle\nonumber\\&+\frac{1}{2}
  \sum_{j\neq a,b}f_{aj}
  \langle{\hat \sigma}_a^z(t)\rangle\langle{\hat \sigma}_j^+
  (t+\tau){\hat \sigma}_b^-(t)\rangle\nonumber,
\end{align}
where we have introduced the approximation $\langle{\hat \sigma}_a^z(t+\tau) {\hat
  \sigma}_j^+(t+\tau){\hat \sigma}_b^-(t)\rangle\approx \langle{\hat
  \sigma}_a^z(t)\rangle\langle{\hat \sigma}_j^+ (t+\tau){\hat
  \sigma}_b^-(t)\rangle$.
Comparisons with exact numerical solutions show that the cumulant
expansion  captures well the steady-state behavior for
inhomogeneous couplings $f({\bf r}_{ab})$, provided the elastic
couplings $g({\bf r}_{ab})$ are sufficiently small. In Fig.~A\ref{subfig:twotime} we compute the pair-wise two-time correlation function, $Z_{a,b}(\tau)\equiv\lim_{t\rightarrow\infty}\langle (\hat
\sigma_a^+ (t+\tau) + \hat\sigma_b^+(t+\tau))(\hat \sigma_a^-(t) +
\hat\sigma_b^-(t))\rangle$, using both the cumulant expansion and the exact solution.  The decay rate  of these correlations,  $Z_{a,b}(\tau) = A e^{-\tau \gamma}$, encodes information about the spectral coherence of the emitted radiation (note that here $\nu = 0$). The result shows that $\Gamma/\gamma$ exhibits the same  dependence on $W/\Gamma$ as $Z_Q$.

\section{{Conditional evolution, entanglement and quantum correlations}}
An individual experimental realization can be considered as a single trajectory, whose evolution can be quite different from the ensemble averaged solution of the master equation. Tracking the evolution of an individual trajectory is equivalent to performing   continuous  measurements that collect the  record  of the emitted photons, for example  homodyne measurements. The conditional evolution of the system subject to continuous measurements
 can be modeled  by  the method of quantum state diffusion~\cite{carmichael,qsdwiseman1}. For a single run the state of the system remains pure, $\hat{\rho}_c=\ket{\psi}\bra{\psi}$,  but the average  over many trials reduces the system into a mixed state and recovers the density matrix obtained from  the master equation.

To probe the entanglement of the dipoles, in Fig.~2 we calculate the average quantum Fisher information for each individual trajectory~\cite{fishersmerzi,Strobel2014}
\begin{eqnarray}
\overline{F}_Q(\hat{\rho}_c)=\frac{1}{3}(F_Q(\hat{\rho}_c;\hat{J}_x)+F_Q(\hat{\rho}_c;\hat{J}_y)+F_Q(\hat{\rho}_c;\hat{J}_z)),\nonumber
\end{eqnarray}
 where  $F_Q(\hat{\rho_c};\hat{\mathcal H})=4(\bra{ \psi} \hat{\mathcal H}^2\ket{\psi}-\bra{\psi}\hat{\mathcal H}\ket{\psi}^2)$, and $\hat{J}_{x,y,z}$ are collective angular momentum operators.

States with zero entanglement can still be nonclassical. Two systems are correlated if they share information with each
other. The total amount of correlation can be quantified by the
quantum mutual information $\mathcal
{I}=\mathcal{S}_A+\mathcal{S}_B-\mathcal{S}_{AB}$, where
$\mathcal{S}_{i}$ is the von Neumann entropy of the subsystem
$i\in \{A,B,AB\}$ ($AB$ is the total system spanned by $A$ and $B$
together) , $\mathcal{S}_i=-\text{Tr}[\hat {\rho}_{i}\text{log}_2{\hat \rho}_{i}]$,
with ${\hat \rho}_i$ the reduced density matrix of the subsystem $i$. A value
varying between 0 and 2 is obtained when $A$ and $B$ are pure states
or maximally correlated respectively. The mutual information  can be separated into
a classical  and a quantum part. The classical part is
$\mathcal{J}_{B|A}=\text{max}\{\mathcal{S}_B-\mathcal{S}_{B|A}\}$.
Here $\mathcal{S}_{B|A}$ is the von Neumann entropy of subsystem $B$
conditioned on the measurement performed on $A$ and $\text{max}$
represents maximum value obtainable  over all local measurements on
$A$. The quantum part, known as the quantum discord,
$\mathcal{D_{B|A}=\mathcal{I}-\mathcal{J}_{B|A}}$, measures the amount
of correlations that exceed the classical part and characterizes the
``quantumness'' of the system~\cite{Modi2012}. A
state with nonzero quantum discord behaves in a way  intrinsically non-classical, since a local measurement performed on
one of its subsystems can disturb the whole system. In
order to calculate  $\mathcal{J}_{B|A}$,
we consider a set of von Neumann measurements
$\hat \Pi_{k=1,2}^A=\frac{1}{2}(1\pm\vec{n}_k\cdot\vec{\sigma}^A)$ with $|\vec{n}_k|^2=1$, made on the subsystem $A$
and minimize the corresponding conditional entropy $\mathcal{S}_{B|A}=\sum_{k=1}^2\text{Tr}[p_k S(\hat \rho_{B|\Pi_k^A})]$, where $p_k=\text{Tr}[\hat \Pi_k^A \hat \rho],\hat\rho_{B|\Pi_k^A}=\text{Tr}_A[\hat \Pi_k^A \hat \rho]/p_k$~\cite{Modi2012}.  In
Fig.~2(e) we calculate the mutual information from
$\mathcal{I}$ and the quantum discord from $\mathcal{D}$ using as subsystems $A$ and $B$  a pair of
dipoles, $a$ and $b$
respectively.  

\bibliographystyle{apsrev}
\bibliography{refsync}
\end{document}